\documentclass[a4paper,10pt]{article}
\usepackage{graphicx,amsfonts}

\begin{document}

\title{Nonstandard mixing in the standard map}



\author{F. Baldovin\thanks{E-mail address: baldovin@cbpf.br},
 C. Tsallis\thanks{E-mail address: tsallis@cbpf.br} and B. Schulze\\
\it{Centro Brasileiro de Pesquisas F\'{\i}sicas}\\
\it{Rua Xavier Sigaud 150,}\\
\it{Urca 22290-180 Rio De Janeiro -- RJ, Brazil } }
\date{August 28, 2001}


\maketitle

\begin{abstract}
The standard map is a paradigmatic one-parameter (noted $a$) two-dimensional
conservative map which displays both chaotic and regular regions. This map
becomes integrable for $a=0$. For $a \ne 0$
it can be numerically shown that the usual,
Boltzmann-Gibbs entropy $S_1(t)=-\sum_{i} p_i(t)\ln{p_i(t)}$
exhibits a {\it linear}
time evolution  whose slope hopefully converges, for very
fine graining, to the Kolmogorov-Sinai entropy.
However, for increasingly small
values of $a$, an increasingly large time interval emerges, {\it before} that
stage, for which {\it linearity} with $t$ is obtained only for
the generalized nonextensive entropic form
$S_q(t)=\frac{1-\sum_{i}[p_i(t)]^{q}
}{q-1}$
with $q = q^*\simeq 0.3$.
This anomalous regime corresponds in some sense
to a power-law (instead of exponential) mixing.
This scenario might explain
why in isolated classical long-range $N$-body Hamiltonians, and depending on
the initial conditions, a metastable state (whose duration diverges with
$1/N\rightarrow 0$) is observed before it crosses over to the BG regime.
\end{abstract}


In his critical remarks about the domain of validity of Boltzmann principle,
Einstein stressed \cite{einstein} that the basis of statistical mechanics lies
on dynamics. Intensive work has recently been done which is consistent
with this standpoint, specifically in situations where anomalous effects
may arise (\cite{tsallis4,costa,lyra,tirnakli,moura,latora1}).
Also, a particularly interesting observation was made in \cite{latora}, where
it was  found a simple connection between the Kolmogorov-Sinai (KS)
entropy  rate (the one that stems from the properties of  mixing of the system)
and the  statistical entropy (the one that originates from a probability
distribution).  It was shown in fact that,
partitionning the phase space
in $W$ cells
and starting from many points within one cell,
the time dependence of the usual,
Boltzmann-Gibbs (BG), statistical entropy
\begin{equation}
 \label{SB}
S_1=-\sum_{i=1}^W p_{i}\ln{p_{i}},
\end{equation}
includes a {\em linear}
stage whose slope coincides with the KS entropy rate, calculated,
via Pesin equality, using the Lyapunov exponents. 
Finally, for one-dimensional dissipative systems at the edge of chaos
(e.g., the logistic map), it was found in \cite{latora1} that the usual
entropy fails to exhibit (with nonvanishing slope)
such a behavior. Instead, the nonextensive entropy
\cite{tsallis} (for a recent review,  see
\cite{tsallis2})
\begin{equation}
\label{ST}
S_q=\frac{1-\sum_{i=1}^W p_{i}^{q} }{q-1},
\end{equation}
succeeds for
a specific value of $q< 1$ ($q^*\simeq 0.2445$ for the logistic map). We remind
that the entropic form (\ref{ST})
reduces to (\ref{SB}) in the limit $q\rightarrow 1$
and that its extremum at equiprobability ($p_i=1/W, \;\;\forall i$) is given by
$S_q=(W^{1-q}-1)/(1-q)$.

In Hamiltonian dynamics, one of the most studied
models is the {\em standard map} (see, for instance, \cite{tabor}), also
referred to as the {\em kicked-rotator} model:
\begin{eqnarray}
\label{std}
x_{t+1}&=&y_t+\frac{a}{2\pi}\sin(2\pi x_t)+x_t~~~\textup{(mod
1)},\nonumber\\ \\
y_{t+1}&=&y_t+\frac{a}{2\pi}\sin(2\pi x_t)~~~~~~~~~\textup{(mod 1)}\nonumber,
\end{eqnarray}
where $a\in\mathbb{R}$. It corresponds to an integrable system when $a=0$, while,
for large-enough values of $a$ (typically
$|a|\geq7$), the map is strongly chaotic
\footnote {We are not interested here in the accelerator-mode islands
that appear at various critical values of $a$ (see, for instance, \cite{zaslavsky}
and references therein).}.
It can be easily verified that, for all values of $a$, this two-dimensional
map is conservative; it has a pair of $(x,y)$-dependent Lyapunov exponents,
which only differ in sign (simplectic structure) and globally vanish
when $a$ vanishes.

In this letter we study the time
evolution of both the extensive ($q=1$) and the nonextensive
($q\neq 1$) statistical entropies of the standard map (\ref{std}), focusing our
attention on small values of the parameter $a$, i.e., on those situations where
the border between the chaotic and the regular regions has a relevant influence.
For decreasingly small values of $|a|$ below say $7$,
one encounters in fact an increasingly rich fractal-like structure,
characteristic of a large class of Hamiltonian systems. Typically, chains of
regular islands corresponding to elliptic points in resonance condition begin
to appear, and between each couple of elliptic points of a chain there is an
hyperbolic point, thus forming another chain of hyperbolic points responsible
for the chaotic areas. Around each
island there is another chain of islands of a higher order and, once again,
another chain of hyperbolic points between the islands (see, for instance,
\cite{henon} for details). This structure is usually referred to as
``islands-around-islands''. If in some sense we think at the
islands-around-islands structure as the edge-of-chaos region of a Hamiltonian
system, we verify that for the standard map its relative area
of influence increases if $a$
approaches $0$.

We know from the theory of chaos that the mixing (herein
we use this word as a synonym of sensitivity to initial
conditions: $\xi(t)\equiv\lim_{\Delta x(0)\to0}\Delta x(t)/\Delta x(0)$, for
a one-dimensional illustration), is typically linear ($\xi-1\propto t$) in the
case of a regular region, and exponential in the case of a strongly chaotic one
($\dot{\xi}=\lambda_1\xi$, hence
$\xi= e^{\lambda_1 t}$, with the Lyapunov exponent $\lambda_1>0$). When
a point falls in the neighborhood of the islands-around-islands structure, it
becomes trapped, a phenomenon called ``stickiness'' (see, for
instance, \cite{zaslavsky}); its motion may be thought in this case as a
succession of ``flights and sticks''. We suggest here, on the basis of the
results that follow, that the presence of the fractal-like structure at the
border between regular and strongly chaotic regions produces a power-law mixing
of the kind (\cite{tsallis4}): \begin{equation}
\xi= e_{q^*}^{\lambda_{q^*} t}\equiv
\left[1+(1-q^*)\lambda_{q^*} t\right]^{\frac{1}{1-q^*}}
\end{equation}
(see \cite{tsallis2} for details on the $q$-exponential function), solution
of the nonlinear equation $\dot{\xi}=\lambda_{q^*}\xi^{q^*}$, ($q^*<1$,
$\lambda_{q^*}>0$).
If we are then to construct a {\em global} solution $\xi
(t)$,
representative of the mixing properities of the {\em whole} phase space,
we must implement some averaging process over regions with different kinds
of mixing.
In this case
we expect that one or even two {\em crossovers} may
happen when time increases, more precisely between the linear and the power-law
mixings, and, later on, between the power-law and the exponential mixings. One way to model
such a behavior is to assume that the differential equation that controls $\xi$
has the form \cite{tsallis3}
$\dot{\xi}=\lambda_1\xi+(\lambda_{q^*}-\lambda_1)\xi^{q^*}$.
The solution
\begin{equation}
\xi(t)=\left[1-\frac{\lambda_{q^*}}{\lambda_1}+\frac{\lambda_{q^*}}{\lambda_1}
e^{(1-q^*)\lambda_1 t}\right]^{\frac{1}{1-q^*}},
\end{equation}
presents, in the case $0<\lambda_1\ll\lambda_{q^*}$, three asymptotical
behaviors, namely (i) linear ($q=0$ regime):
$\xi\sim 1+\lambda_{q^*}t$, for $0\leq t\ll
t_{cross1}\equiv\frac{1}{(1-q^*)\lambda_{q^*}}$;
(ii) power-law ($q=q^*$ regime):
$\xi\sim\left[(1-q^*)\lambda_{q^*}\right]^{\frac{1}{1-q^*}}
t^{\frac{1}{1-q^*} } $, for $t_{cross1}\ll t\ll t_{cross2}\equiv
\frac{1}{(1-q^*)\lambda_1}$; (iii) exponential ($q=1$ regime):
$\xi\sim (\frac{\lambda_q}{\lambda_1})^{\frac{1}{1-q}}e^{\lambda_1 t}$, for
$t\gg t_{cross2}$.

With this picture in mind, we explore now numerically the behavior of the
entropy with time. We start introducing a coarse-graining
partition of the mapping phase space by dividing it in $W$ cells of equal
size, and we set many copies of the system ($M$ points) in a
far-from-equilibrium situation putting all the $M$ points randomly or
uniformly distributed inside a single cell.
The occupation number $M_i$ of
each cell $i$ ($\sum_{i=1}^W M_i=M$) provides a probability distribution
$p_i\equiv M_i/M$, hence an entropy value. Using then
the dynamic equations (\ref{std}), at each step the points spread in the
phase space causing the entropy value to change. In order to obtain a numerically
consistent definition of the probabilities $\{p_i\}$ when the system spreads at
its maximum, we set $M>10\; W_{max}$, where $W_{max}\leq W$ is the maximum
number of cells that the initial data occupy for the steps we are
considering. Finally, to extract a {\em global} quantity on the $(x,y)$ phase
space, we
repeat the calculation setting the cell that contains the initial points in
different positions chosen randomly all over the {\em whole} unit square, and we
take an average over all the different histories thus obtained.
We stress here that in taking this average all over the
phase space we are consistent with a Gibbsian point of view, thus
obtaining a statistical description of how the
system approaches equilibrium.
The result of this
analysis for fixed $a$
is then a single curve of the entropy versus time, for each entropic
form $S_q$.

When the mixing is exponential and we use $q= 1$, we obtain the behavior
illustrated in \cite{latora}: a {\em linear} increase just before the saturation due to
the finiteness of $W$, with a slope approaching, for $W\to \infty$,
the positive Lyapunov exponent.
If we vary $q$, we have that for $q<1$
$(q>1)$ the entropy grows with positive (negative) concavity before
saturation.
The exponential mixing is telling us that there is an unique value of $q$,
namely $q=1$,
which produces a {\em linear} growth in the entropy.
We schematize this analysis in Fig. 1.
Nevertheless, if some phenomenon like the trapping one described in
\cite{zaslavsky} causes the mixing to slow down, thus becoming power-law, the same
analysis exhibits that the only value of $q$ for which the entropy
displays a {\em linear} stage with time is $ q=q^*$, with $q^*<1$
(see, for example, \cite{latora1}).
Finally, if the mixing is linear, the specific value of $q$
corresponding to a linear stage is $ q=0$.

In the case of the standard map the situation is more intricated because
we have a coexistence of regions with
different mixing properties. Fixing, for
example, $a=5$, the phase space presents, macroscopically, two
islands and a connected chaotic sea. Depending on where we set the initial data,
we may face a strongly chaotic, or a regular, or even an
islands-around-islands region. Our averaging procedure does not
privilege any of these regions and allows the predominant one to emerge. The
key point is to perform many
different histories so that the average stabilizes on a
definite curve. The resulting entropy curves are then sensible to two
different phenomena: the characteristics of the mixing of each area (represented
by the Lyapunov exponent $\lambda_1$, or its generalization
$\lambda_{q^*}$ in the case of a power-law mixing)
and the relative extension of each area with respect to the whole phase space
(for this reason, in the following we indicate the slopes of the linear phases
of $S_q(t)$ with $\widetilde\lambda_q$).
If the phase space exhibits
a predominance of power-law mixing region associated with a generalized
Lyapunov exponent $\lambda_{q^*}$, with $q^*<1$,
we expect, for time not too large,
a {\em linear} growth for the entropy $S_{q^*}$.
Waiting enough time,
a crossover to the exponential mixing would occur, due to the rapidity of
the exponential growth with respect to the power-law one,
regardeless of how small is the exponential mixing region.
After the crossover, the linearity with $t$ is obtained for $S_1$
(and not for $S_{q^*}$).
We represent this more complex situation in Fig. 2, in the limit
$W\to\infty$.

Let us present now our computational results for typical values of $a$.
Fig. 3 represents the case $a=5$ and exhibits the predominance of
the exponential-mixing. After the first step, that
here and in what follows we consider
as a transient, only the BG entropy $S_1(t)$ has a
linear stage before the effects of saturation,
while $S_q$ has positive concavity for $q=0.7$
and negative concavity for $q=1.3$. For $q=1$, our estimation of the slope
$\widetilde\lambda_1(5)=0.96$ is different from
that found in \cite{latora} because here we average all over
the phase space, {\em including the islands}.

We notice that the scale that each entropic form
covers, varies widely. To perform a linearity analysis between different
entropy curves we
use the following quite sensitive method. First, we
fit the relevant piece of each curve (between $t=t_1$ and $t=t_2$) with a
polynomial of the second degree: $At^2+Bt+C$. Then, we define the linearity
coefficient $\rho\equiv A/(At_2^2+Bt_2-At_1^2-Bt_1)$; $\rho$ tests
the value of the second derivative of the fitting polynomial, including a scale
renormalization. Obviously, $\rho=0$ corresponds to linearity, while $\rho>0$
and $\rho<0$ correspond to positive and negative curvatures respectively. An
important remark is that the effect of saturation for a finite value of $W$
decreases the curvature of an entropy curve with respect
to the case $W\to\infty$. This in turns introduces a systematic underestimation
of $\rho$, which increases when $t_1$ and $t_2$ increase.

In Fig. 4 we show how, decreasing $a$, the first part of the $S_1(t)$ becomes concave.
After the crossover the curves exhibit once again a linear increase.
The slope of the linear stage
becomes smaller, which suggests that
the Lyapunov exponent $\lambda_1$ is globally going to $0$.
In correspondence with the negative concavity of the first part of $S_1(t)$, $S_{0.3}(t)$
displays a linear increase (Fig. 5), with a slope that does not decrease as fast as
the one of $S_1(t)$ (see the inset of Fig. 4).
After the crossover, $S_{0.3}(t)$ becomes convex.
We notice also that the crossover time $t_{cross}$ increases when $a$ decreases
(inset of Fig. 5). It is the same linearity analysis performed in Fig. 3(d)
which leads to
$q^*\simeq 0.3$.

Summarizing, we have studied the production of entropy for a well known
low-dimensional conservative map controlled by the parameter $a$.
It appears that, for fixed $a\neq 0$ and in the limit $t \rightarrow \infty$
(to be in all cases taken {\it after} the $W \rightarrow \infty$ limit),
it is the standard, BG entropy $S_1$, which is
associated with a {\it finite} entropy production. However, during the
time interval preceeding this extreme limit, an increasingly large
interval emerges for which it is $S_{q^*}$, with $q^* \simeq 0.3$, which
is associated with a {\it finite} entropy production. Consistently,
our results suggest that,
in the $lim_{t \rightarrow \infty}\;lim_{a \rightarrow 0}$
ordering, the only regime in fact observed is that corresponding
to $q^* \simeq 0.3$. This fact provides a very appealing scenario
for (meta) equilibrium thermostatistics in long-range Hamiltonians
such as that considered in \cite{latora2}. For these systems, a
longstanding metastable state can exist (preceeding the BG one)
whose duration diverges with $N \rightarrow \infty$, and whose distribution
of velocities is {\it not} Maxwellian, but rather a power-law. The role
played by $a$ in our present simple map
and that played by $1/N$ in such long-range-interacting many-body systems,
might be very similar, thus providing a
dynamical basis for nonextensive statistical mechanics
\cite{tsallis}.

\section{Acknowledgments}    
We acknowledge E.G.D. Cohen for stressing our attention on Einstein's 1910
paper. We have benefitted from partial support by PRONEX, CNPq, CAPES and
FAPERJ (Brazilian agencies).


\newpage
\begin{figure}
\begin{center}
\includegraphics{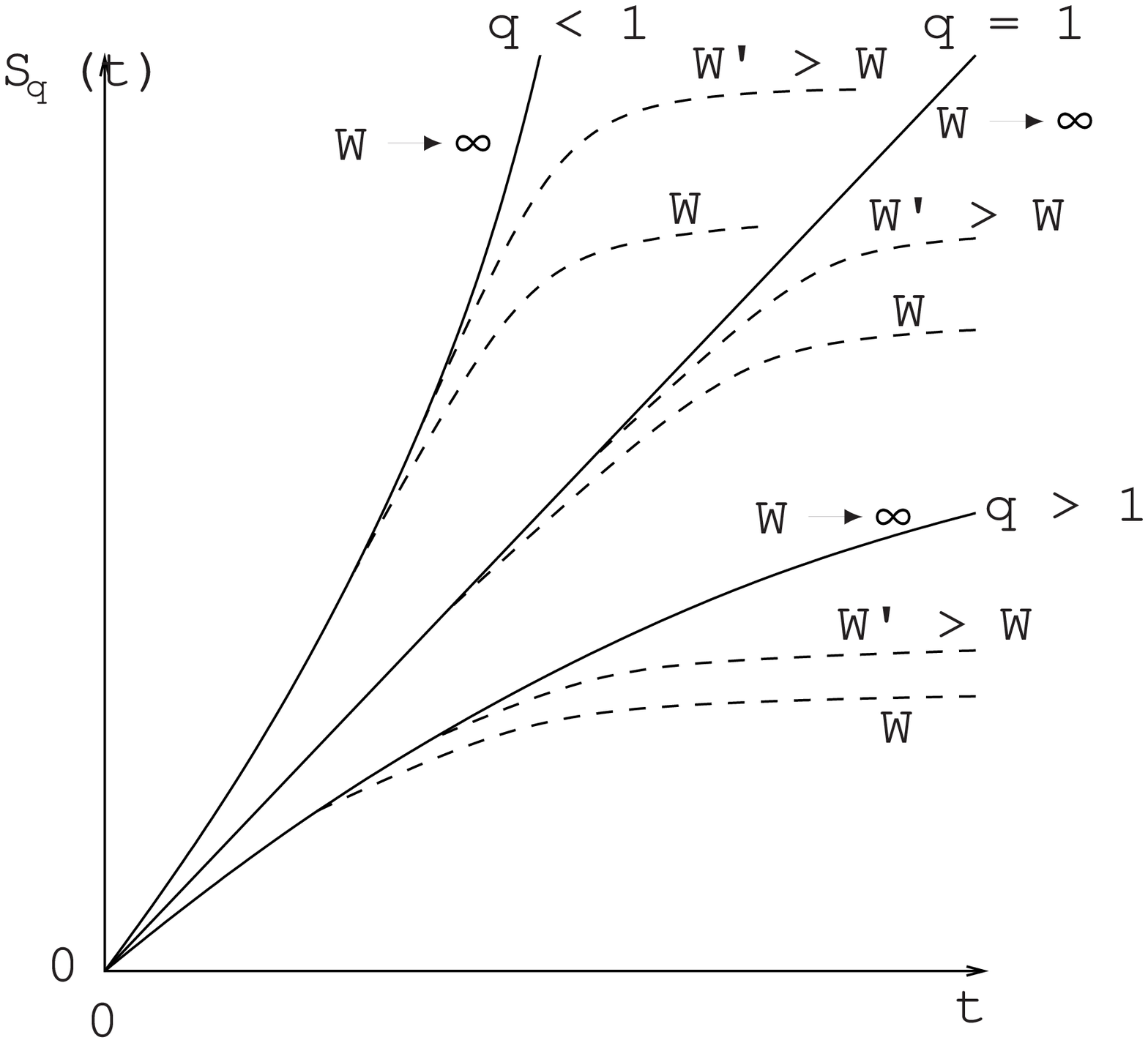}
\end{center}
\caption{\small Schematic representation of $S_q(t)$ for different
values of $q$, in the case of exponential mixing.
Only for $q=1$ we have a {\em finite} entropy production
($\lim_{t\to\infty}S_q(t)/t$).}
\end{figure}

\newpage
\begin{figure}
\begin{center}
\includegraphics{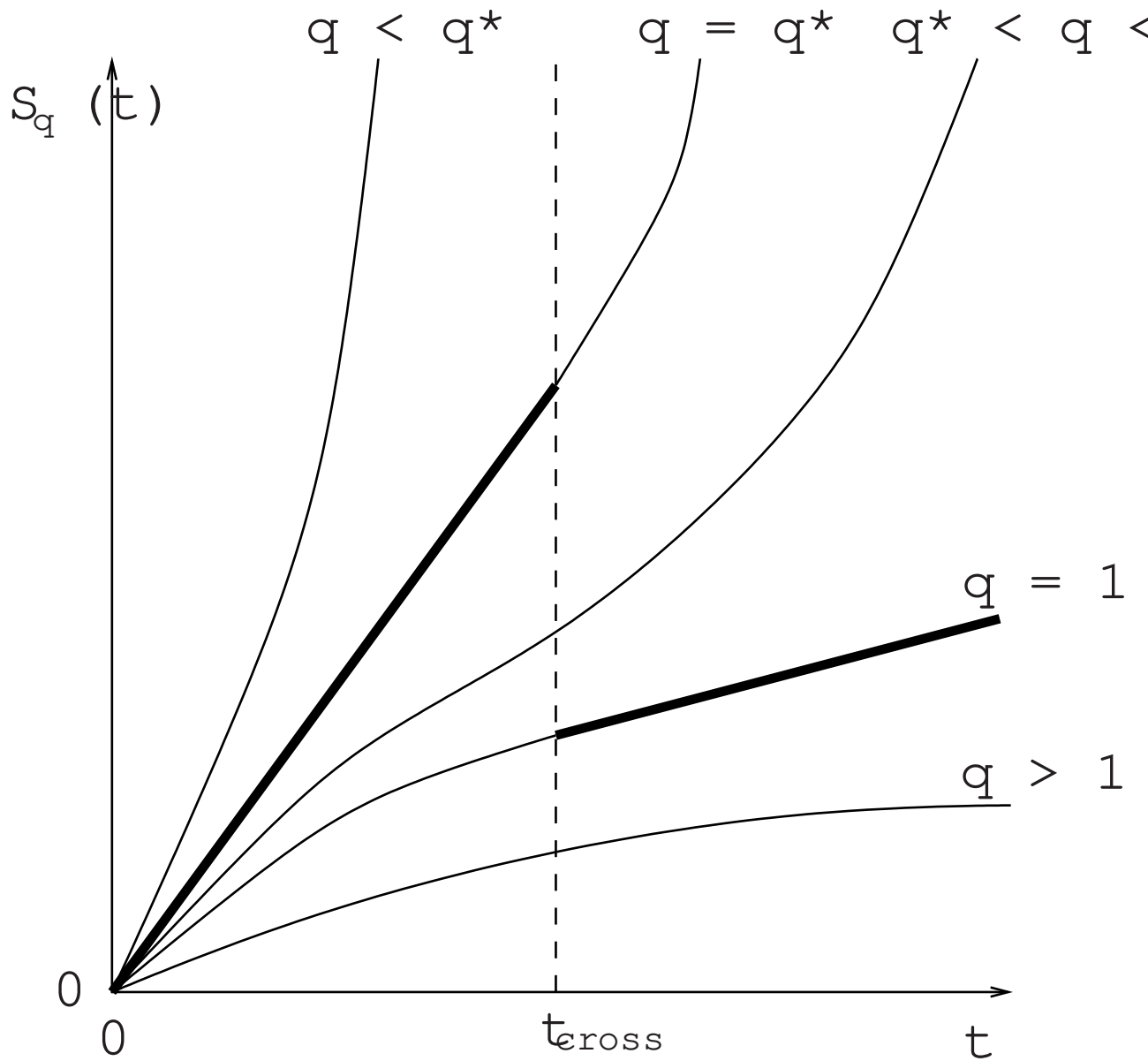}
\end{center}
\caption{\small Schematic representation of $S_q(t)$
for different values of $q$, in the limit $W\to\infty$ and
when there is coexistence between power-law ($q^*<1$) and
exponential mixings. The thicker lines emphasize the linear regions.}
\end{figure}

\newpage
\begin{figure}
\begin{center}
\includegraphics{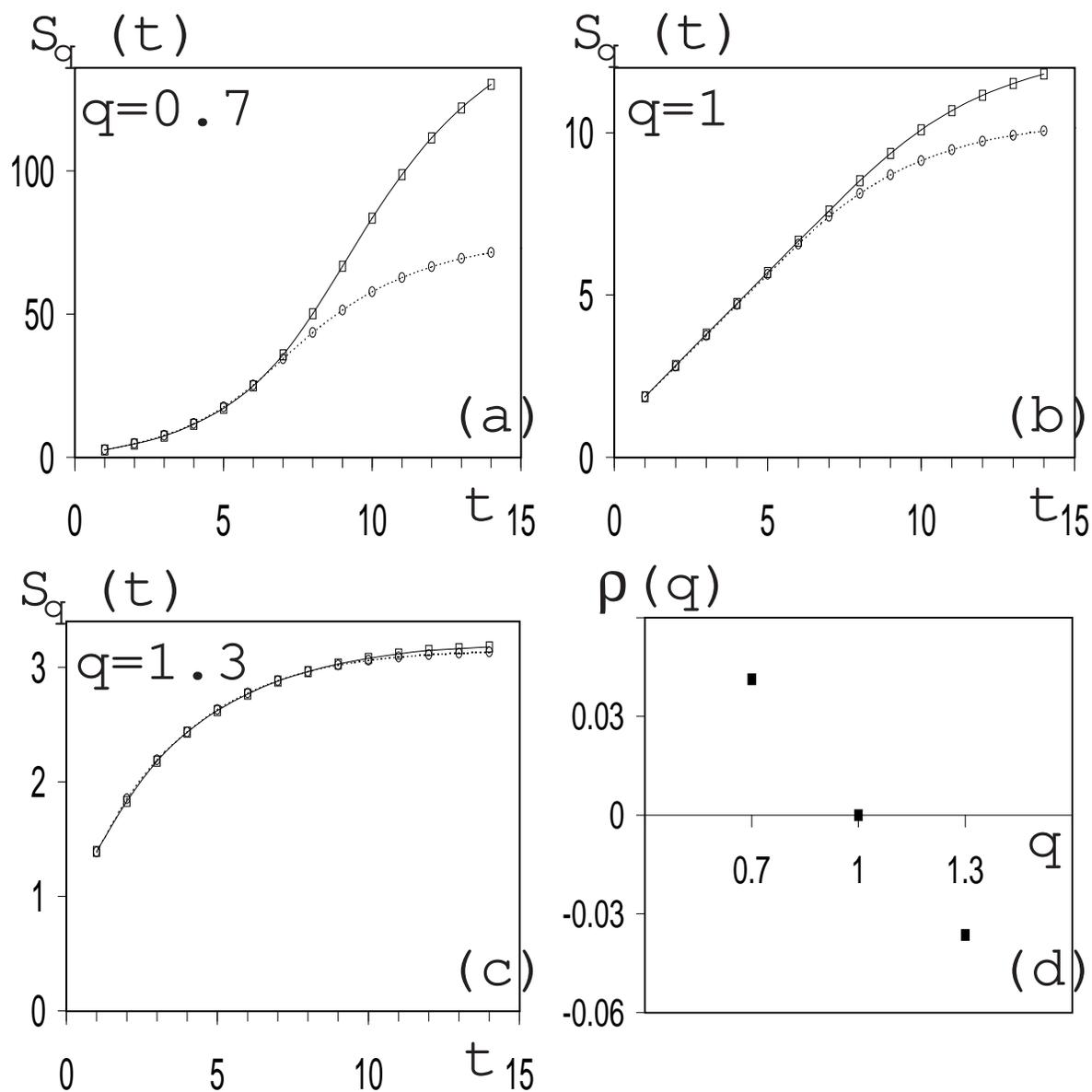}
\end{center}
\caption{\small Typical $S_q(t)$ for $a=5$ (a,b,c).
$1000$ histories were averaged; squares correspond to $M/10=W=708\times 708$;
circles correspond to $M/10=W=224\times 224$.
The lines are guides to the eye.
(d) Linearity analysis $\rho(q)$ for the squares, with
$t_1=2$ and $t_2=6$ (see text).
The slope of $S_1(t)$ between $t_1=2$ and $t_2=6$ is
$\widetilde\lambda_1(5)=0.96$.}
\end{figure}

\newpage

\begin{figure}
\begin{center}
\includegraphics{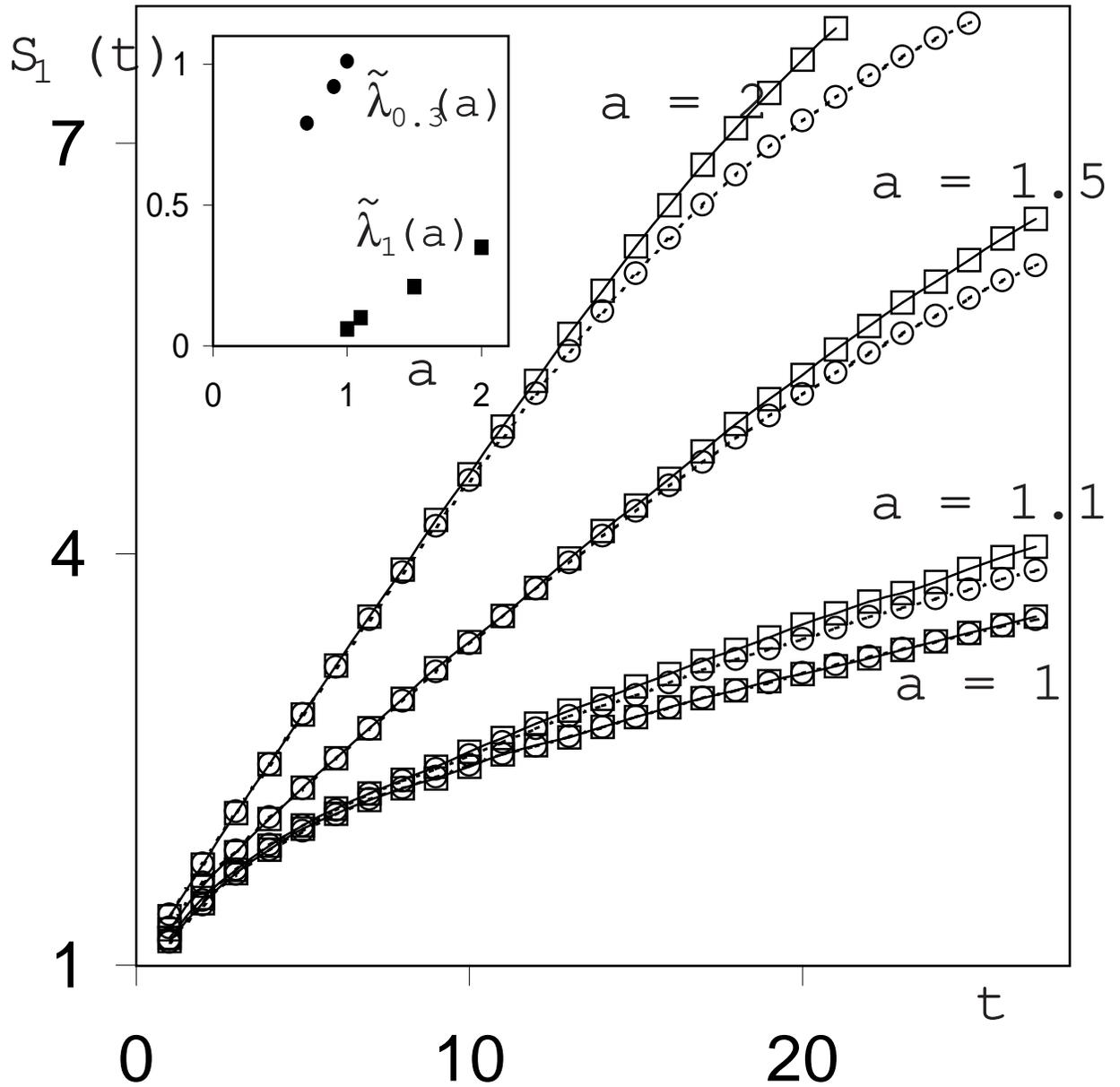}
\end{center}
\caption{\small $S_1(t)$ for typical values of $a$
($5000$ to $7000$ histories were averaged); squares correspond to $M=W=1000\times 1000$
for $a=2,1.5,1.1$
($M=W=2236\times 2236$ for $a=1$);
circles correspond to $M=W=448\times 448$ for $a=2,1.5,1.1$
($M=W=1000\times 1000$ for $a=1$).
The lines are guides to the eye.
Inset: slopes of $S_1(t)$ and $S_{0.3}(t)$ (see also Fig. 5)
in their linear stage.}
\end{figure}

\newpage
\begin{figure}
\begin{center}
\includegraphics{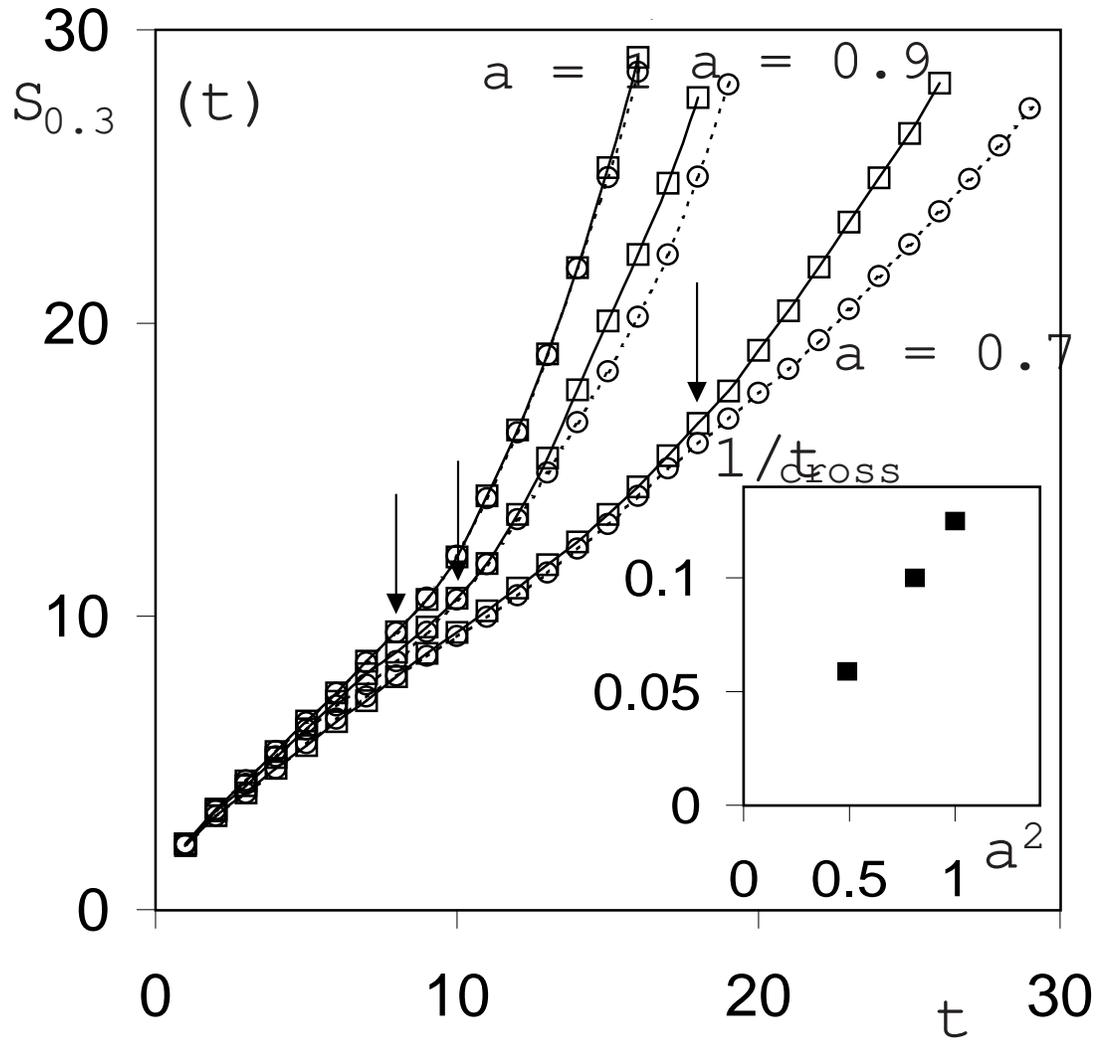}
\end{center}
\caption{\small $S_{0.3}(t)$ for typical values of $a$
($5000$ to $16000$ histories were averaged);
squares correspond to $M=W=2236\times 2236$;
circles correspond to $M=W=1000\times 1000$.
The lines are guides to the eye.
The arrows indicate the crossover time $t_{cross}$ for the corresponding values
of $a$, as represented in the inset.}
\end{figure}

\end{document}